\documentclass[aps,prl, reprint, superscriptaddress]{revtex4-1}
\usepackage{graphicx}

\begin{document}
\title{Light focusing in the Anderson Regime}

%
%
%
%
%
%

\author{Marco Leonetti}
\affiliation{IPCF-CNR c/o Physics Department, University of Rome La Sapienza, P.le A. Moro 2, 00185, Rome, Italy}
\affiliation{Center for Life Nano Science@Sapienza, Istituto Italiano di Tecnologia, Viale Regina Elena, 291 – 00161 Roma (RM) – Italia}

\email[Corresponding Author: ]{marco.leonetti@roma1.infn.it}

\author{Salman Karbasi}%
\affiliation{Department of Electrical Engineering and Computer Science
University of Wisconsin-Milwaukee
Milwaukee, WI 53201, USA}

\author{Arash Mafi}%
\affiliation{Department of Electrical Engineering and Computer Science
University of Wisconsin-Milwaukee
Milwaukee, WI 53201, USA}

\author{Claudio Conti}
\affiliation{Dep. Physics University Sapienza, P.le Aldo Moro 5, I-00185, Roma Italy}

\begin{abstract}

\end{abstract}

\date{\today}
\maketitle
\textbf{Anderson localization is a regime in which diffusion is inhibited and waves (also electromagnetic waves) get localized. Here we exploit adaptive optics to achieve focusing in disordered optical fibers in the Anderson regime. By wavefront shaping and optimization, we observe the generation of a propagation invariant beam, where light is trapped transversally by disorder, and show that Anderson localizations can be also excited by extended speckled beams. We demonstrate that disordered fibers allow a more efficient focusing action with respect to standard fibers in a way independent of their length, because of the propagation invariant features and cooperative action of transverse localizations.}

Adaptive techniques have turned around optics allowing not only to correct aberrations in the image formation but also to focus light beams through curtains of dielectric scatterers\cite{vellekoop2010exploiting, vellekoop2008universal, popoff2010measuring} by employing the most transmitting modes\cite{kim2012maximal}. The focusing may be achieved by wavefront-shaping, using spatial light modulators (SLMs) and applying a specific phase distribution to the input beam to correct the random delay imposed by the diffusive propagation. Focusing by wavefront shaping has opened the way to many novel applications \cite{katz2012looking,akbulut2011focusing, van2011scattering} as it can be realized in any disordered structure, even if, so far, it has been studied only within the diffusive regime, in the absence of mechanisms of wave-localization.
It is well accepted that if the strength of disorder increases beyond a critical value, a transition called Anderson localization\cite{ShengBook,lagendijk2009fifty,anderson1958absence} takes place. In the proximity of this regime, there is a drastic reduction of diffusion, and ultimately an absence of transport. Being originated by interference, Anderson localization is common to all kind of waves and has been demonstrated for matter waves\cite{roati2008anderson}, sound\cite{hu2008localization} and entangled photons\cite{crespi2013anderson}. For light it is difficult to achieve localization
in three dimensions (3D) \cite{sperling2012direct} because the scattering strength has to be strong enough to satisfy the Ioffe$-$Regel criterion,  while absorption has to be negligible\cite{anderson1985question,wiersma1997localization,PhysRevLett.58.2486,storzer2006observation}.
On the contrary, two dimensional (2D), or {\it transverse}, localization \cite{de1989transverse} is always obtained in sufficiently large samples and has various analogies with the focusing through adaptive processes : they are both coherent phenomena and allow to trap light in a tiny spot. Transverse localization \cite{de1989transverse} occurs in systems disordered in the plane perpendicular to the direction of propagation: it has been demonstrated in arrays \cite{pertsch2004nonlinearity}, in optical lattices  \cite{schwartz2007transport}, and in plastic \cite{karbasi2012observation} and glass \cite{karbasi2012transverse} fibers.

\begin{figure}
\includegraphics[width=8 cm]{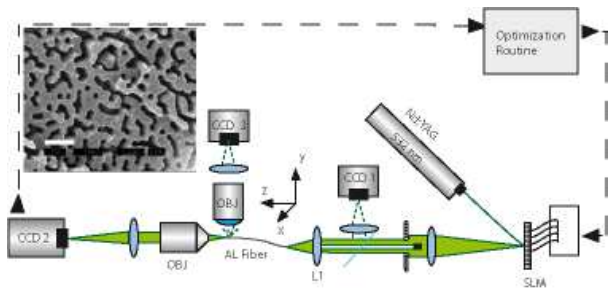}
\caption{ Light from a solid state ND:YAG laser is reflected by an SLM
in phase only configuration obtained by blocking the first diffracted order by a beam stop. Modulated light is coupled by the lens L1 onto the fiber. Input and output faces of the fiber are imaged by microscopy objectives (OBJ)  on CCD 1 and CCD 2, respectively. A computer-aided feedback control the SLM. The lateral side of the fiber tip is imaged on CCD 3. The inset shows a scanning electron microscope image of the fiber tip, in which darker regions correspond to PMMA, and white scale is 4 $\mu$m. \label{setup}}
\end{figure}

\begin{figure}
\includegraphics[width=8 cm]{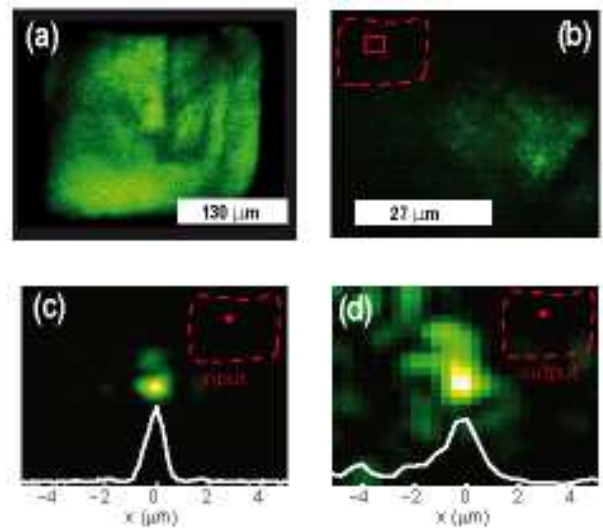}
\caption{ (a) laser at the fiber output (side $250$~$\mu$m)
(b) as in (a) averaged over $500$ random input, and zoomed on the position indicated in the red sketches; (c) laser at the fiber input face in the single mode injection ($0.7$~$\mu$m input spot); (d) output of the fiber corresponding to the input condition in (c). The location of the hotspots in the fiber is sketched in the insets.
\label{Fiberimages}}
\end{figure}

The role of transverse localization in speckle focusing has never been previously investigated. Here we study the interplay between the focusing process and Anderson localization and demonstrate that the absence of diffusion cooperates with the optimization protocol  to improve focusing effectiveness. We use an optical fiber (without cladding) with transverse disorder and binary refractive index modulation  \cite{lagendijk2009fifty,SalmanOL,SalmanOPEX}, (see inset of figure \ref{setup}). The index contrast of the order of  0.1 results from the difference between between Polystyrene (PS, refractive index $n=1.59$) and polymethylmethacrylate (PMMA $n=1.49$). We estimate the value of losses in 0.5 Db/cm. This value will be further decreased of at least two orders of magnitude exploiting advanced fabrication techniques. Moreover a relevant part of the losses are due to the input and output coupling and can be  reduced by improving the fiber cutting procedure.
The disordered fiber has been fabricated by melding 40,000 strands of polystyrene  and 40,000 strands of polymethyl methacrylate. The mixture of strands was fused together and redrawn to a square shaped fiber with a lateral size $W=250~\mu$m \cite{karbasi2012observation}. Samples have length $Z_L$ between 1.2 and 8~cm. Light generated by a continuous-wave (CW) laser with wavelength $\lambda=532$~nm (vertical polarization)  is injected in the system after being modulated by an SLM in the phase only configuration (experimental setup shown in Fig.~\ref{setup}). Fig.~\ref{Fiberimages}a shows the fiber output when the input is completely illuminated (horizontal polarization is retrieved to eliminate ballistic light). The focusing effect is found to be independent of the input polarization. Such a structure supports strong localization (as already demonstrated in \cite{karbasi2012observation,SalmanOL}), having a refractive index mismatch three orders of magnitude larger than in the seminal experiment of \cite{schwartz2007transport}, and also allows image transmission \cite{karbasi2013image}.

\begin{figure}
\includegraphics[width=8 cm]{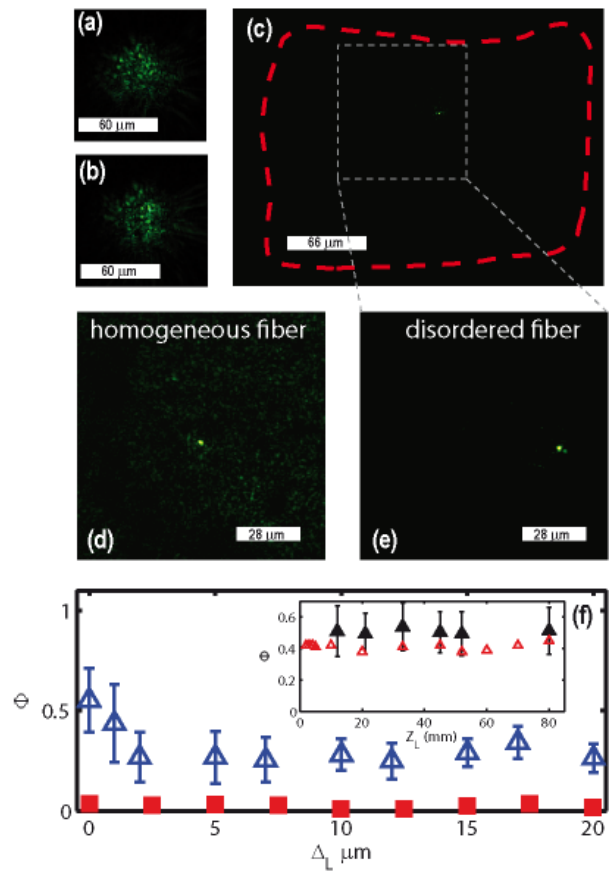}
\caption{Input spot on the fiber before (a) and after (b) the optimization procedure by CCD~1. Beam size is $60$~$\mu$m.
(c) Fiber output after the optimization procedure by CCD~2.
(d) Fiber output with magnification higher than (c) for a PMMA homogeneous fiber; (e) as in (d) for a ALF.
(f) Focus efficiency $\Phi$ as a function of the distance $\Delta L$ from an output hotspot for the homogeneous PMMA fiber (full squares) and for the ALF (open circles).
Error bars are given by the statistics on five samples, and are smaller than symbols in the homogeneous fiber.
The inset shows $\Phi$ as a function of the fiber length $Z_L$, measured experimentally (full triangles) and numerically calculated (open triangles, $V_0=100$).
 \label{Optimization}}
\end{figure}

Independently of the spatial shape of the input beam, some particular hotspots at fixed positions in the fiber output are observed, as in  Fig.~\ref{Fiberimages}b, which shows the transmitted intensity from the output face of the fiber. Fig.~\ref{Fiberimages}b is obtained by averaging $500$ random input configurations realized by a random SLM phase-mask  producing a $60~\mu$m speckled spot at the fiber entrance (see also Fig.~\ref{Optimization}a,b). The presence of hotspots appearing at fixed positions independently of the input mask is connected with the presence of extremely efficient transport channels\cite{kim2012maximal}. The high transmitting channels are Anderson modes of the fibers, i.e., transverse localizations (TL) that retain a fixed transverse profile along propagation. This is demonstrated by injecting light by a long working-distance objective (numerical aperture $0.8$, resulting focused spot size 0.7~$\mu$m), which feeds selectively a single mode. In this configuration the input and the output spots are similar in size (Fig.~\ref{Fiberimages}c,d), and appear at the same transverse location in the fiber. This confirms that the excited mode is an eigenmode of the fiber, and its shape and position are not affected by the propagation.

Having located such efficient channels, we applied a feedback algorithm to improve the intensity in one of the observed hotsposts. Our approach is a standard one, very similar to that described in\cite{vellekoop2008phase}: a random phase shift is applied to a segment of the SLM, which is divided in a 23$\times$23 matrix composed by 529 segments. The CCD~1 in Fig.~\ref{setup} grabs an image of the output and the algorithm retains the change in the phase mask  only if the intensity in the hotspot increases,  otherwise the previous phase shift is restored. At the end of the optimization procedure (after 529 steps) more than one half of the fiber output has been channeled into the target point, i.e., into a four micrometer area (fiber lateral side  $W=250$~$\mu$m). The input speckled beam ($60$~$\mu$m waist, obtained by a $50$~mm focal lens), is shown in Fig.~\ref{Optimization}a (before the optimization) and Fig.~\ref{Optimization}b (at the end of the optimization). Fig.~\ref{Optimization}c shows the image of the output focused beam, and demonstrates that, in the Anderson regime, one half of the transmitted energy is focused in a $2$~$\mu$m$^2$ squared area centered at the target, the ratio of the intensity at focus with respect to the average background is $10^4$.

For comparison, we repeated the experiment by using an homogeneous disorder-free standard fiber and found very different results: the optimized focused spot contains only $1\%$ of the total transmitted intensity, and the ratio of the intensity at focus with respect to the average background is of the order of 50; this is a result comparable with the state of the art in multimode fibers\cite{di2011hologram, Cizmar:11,vcivzmar2012exploiting}. The same comparison cannot be done with photonic crystal fibers, which typically support few modes thus making impossible the optimization protocol. The approach proposed here has several advantages also if compared with fiber bundles:  the absence of an alignment requiring mechanical movement of the optics, and a larger set of possible outputs. All the possible positions of the output facet may be targeted also simultaneously generating multiple foci.
Images of the focus for the homogeneous PMMA   fiber and for the Anderson fiber (ALF) are shown in panels \ref{Optimization}d and \ref{Optimization}e, respectively.

\begin{figure}
\includegraphics[width=8 cm]{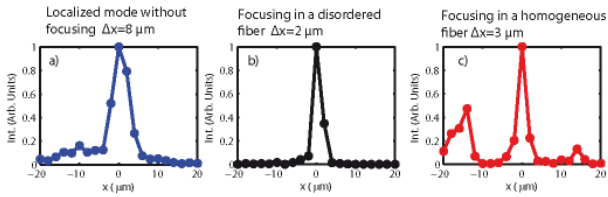}
\caption{ Intensity profile of the focus. (a) Profile of a localized mode without optimization. (b) As in  (a) after the optimization procedure. (c) A focused mode in an homogeneous fiber; note the background.
\label{profiles}}
\end{figure}

A further comparison between various cases is in Fig. \ref{profiles} in which the profiles of the focus in various configurations are reported.

At variance with the standard PMMA fiber, the presence of TL introduces a strong dishomogeinity of the response in the transverse direction. We performed a set of measurements (shown in Fig. \ref{Optimization}f) with variable distance $\Delta L$ from an high transmitting transport channel, determined by the position of the output hotspots as described above.
Specifically, instead of maximizing the intensity in correspondence of a chosen hotspot, we maximized the intensity at a distance $\Delta L$ from it,  and measured the focusing efficiency, defined as the ratio $\Phi$ between the power channeled in this shifted position (at the end of the optimization procedure) and the total output power from the fiber. The result in Fig.~\ref{Optimization}f shows that $\Phi$ rapidly decreases when moving far away from the hotspot, which hence represents the most efficient position of the focusing (and corresponds to a TL). On the contrary, the homogeneous fiber (red squares  in Fig.~\ref{Optimization}f) shows no significant variation of $\Phi$ in terms of the target position $\Delta L$.

This difference is ascribed to the presence of TL. To demonstrate that the propagation invariant TL is the leading mechanism to concentrate light at the target point, we repeated the focusing experiments with fibers with varying length $Z_L$. The inset in Fig.\ref{Optimization}f shows that $\Phi$ is nearly indepedent of $Z_L$, as is also confirmed by results from the numerical simulation described below. The most effective focusing in the disordered case is due to the presence of the Anderson localization that inhibits light diffusion and practically eliminates the background speckle pattern. We stress that the TL focusing action is concomitant with the standard focusing action, which involves all the modes of the fibers and results in an enhancement of the intensity at the focus.

To further confirm this scenario, we resorted to the numerical simulation of the paraxial equation approximating the propagation of an optical field $A$ (with $I=|A|^2$ the beam intensity) in the ALF \cite{ArnaudBook}. In dimensionless units this equation reads as
\begin{equation}
i\partial_z a+\nabla^2_\perp a-V(x,y)a=0,
\label{norm1}
\end{equation}
with $V=-2k^2 W^2 \Delta n/n_0$, a=A$/\sqrt{I_0}$, being  $I_0$ a reference intensity,
$k=2\pi n_0/\lambda$ and $n_0$ the average refractive index. The transverse disorder is given by a term representing the random fluctuations of the refractive index in the transverse plane $\Delta n=\Delta n(X,Y)$, and $z=Z/(2kW^2)$ and $(x,y)=(X/W,Y/W)$:

Eq.~(\ref{norm1}) is an approximate scalar model for beam propagation in a high index contrast ALF. However, while fully vectorial calculations allow determining the supported mode profiles, the simulation of the optimization procedure and beam propagation is computationally prohibitive beyond the scalar approximation in Eq.~(\ref{norm1}).

The transverse bound states (2D Anderson localizations) in this simplified model, are given by
$a=\varphi_n(x,y)\exp(i E_n z)$
with
\begin{equation}
-\nabla_\perp^2 \varphi_n+V \varphi_n=E_n \varphi_n\text{.}
\label{modes1}
\end{equation}
\begin{figure}
\includegraphics[width=8 cm]{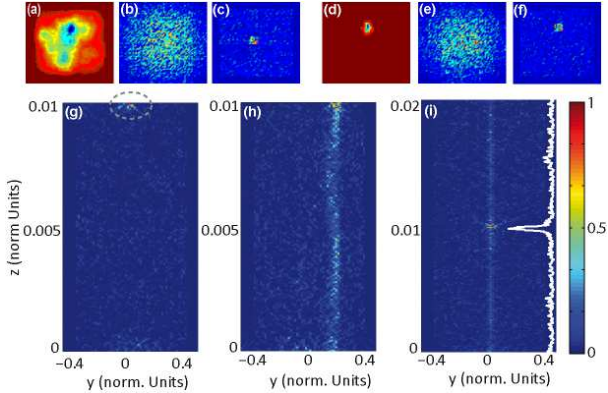}
\caption{
Results from simulations of Eq.(\ref{norm1}) in the low index contrast ($V_0=10$) configuration: (a) ground state, (b) input
after optimization, (c) output after optimization, bottom panel (g): intensity distribution in the section
of fiber in correspondence of the output spot along propagation direction; note that the focusing occurrs in the last part of propagation as indicated by the dashed circle.
\\
Results after Eq.(\ref{norm1}) for high index contrast ($V_0=100$):
(d) ground state, (e) input
after optimization, (f) output after optimization, bottom panel (h): intensity distribution in the section
of fiber in correspondence of the output spot along propagation direction. Note that the localization
since the beginning of propagation and further enhanced at the end; (i) as in (h) for a doubled propagation length and with the optimization target located in the middle of the fiber. The white line is the intensity profile in correspondence of the focus Vs. Z.
\label{fig_claudio1}}
\end{figure}

Because of the mentioned computational limitations, we approximate the disorder distribution by Gaussian random potential, such that $<V(x,y)V(x',y')>=V_0^2\delta(x-x')\delta(y-y')$.
 We remark that the distribution of the disorder is coarse-grained in the numerical simulations by the
adopted discretization, and by retaining as independent Gaussian variables the noise values in different grid points.
We verified that the discretization does not affect the reported results by increasing the number of grid points in
the numerical simulations.
We also considered binary random potentials (not reported) with results similar to what follows.
The Gaussian potential is also included in a rectangular well to account for the interface between the fiber and air.
The strength of the potential $V_0$ is determined by the refractive index contrast: the index jump between PS and PMMA is $\Delta n=0.1$ as the latter varies on a spatial scale $\Delta_D=1~\mu$m of the order of $\lambda$, we have $V_0=2 k^2 W^2 \Delta_D \Delta n/n_0$, which gives $V_0\cong 100$ for our fiber; for comparison we take $V_0=10$ for the weak disorder cases. In our trials we varied $V_0$ in a range of two decades and found no qualitative changes with respect to the representative cases reported in the following; we stress that the considered propagation distances correspond to those in our experiments (in our normalized units $z=1$ corresponds to $Z_L\cong1$~m).

We first determine the 2D eigenmodes in our simplified model from (\ref{modes1}), as given in Fig.~\ref{fig_claudio1}a (Fig.~\ref{fig_claudio1}d)
for the low (high) index contrast case. The mode with the stronger localization (ground state) is chosen as the target for the optimization.
By solving equation (\ref{norm1}), the field profile at each point (x, y and z ) is calculated; the input condition is generated
as done in the experiments and results in speckled beam are shown in Fig.\ref{fig_claudio1}b (Fig \ref{fig_claudio1}e) for the low (high) index contrast.
After a random modification of the phase at the input, the solution of (\ref{norm1}) is calculated, and the change is retained if the intensity at the target point increases.
When the index contrast is very small (i.e., far from the strong localization condition), the focusing appears only at the very end of the fiber
(figure \ref{fig_claudio1}g), that is  at a well defined z, as it occurs in 3D with standard materials\cite{vellekoop2010exploiting}, or in multimode optical fibers \cite{di2011hologram}. When the degree of localization is strong  (figure \ref{fig_claudio1}h), the focus appears in correspondence of a TL, propagating along the direction z, being  enhanced at the fiber tip.
\begin{figure}
\includegraphics[width=8 cm]{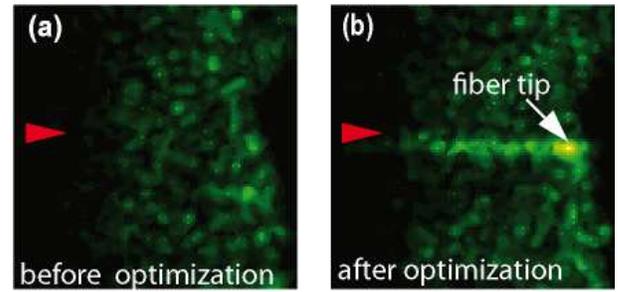}
\caption{(a) light scattered from the side of the fiber in correspondence of the exit tip before the optimization process; (b) as in (a) after the optimization process. The side of the panels is $160$~$\mu$m. \label{Section}}
\end{figure}
In Fig. 4i we show the intensity profile after an optimization for a target located in the middle of fiber, and found a pronounced intensity peak in the focus. This shows that most of the energy is carried by the TL located at the target, and that the other modes interfere constructively to enhance the local optimized intensity. As in experiments, the numerical simulations furnish an efficiency $\Phi$ independent of the fiber length (see inset of Fig.~\ref{Optimization}f). The numerically calculated $\Phi$ is found to be slightly smaller than the experimentally measured value due to the limited resolution of the simulations, where we also considered very small fiber lengths not accessible in the experiments ($Z_L<12$~mm in Fig.~\ref{Optimization}f).

We experimentally verified the existence of such propagating mode appearing together with the  focus in the ALF by measuring the light scattered at one side of the fiber, by using a microscope and CCD~3 (see Fig. \ref{setup}). Results are reported in Fig.~\ref{Section} where the arrow indicates the direction of light propagation. Fig.~\ref{Section}a shows the image of the ALF side before the focusing-optimization process, and Fig.~\ref{Section}b shows the same fiber section at the end of the focusing-optimization process. In the latter case, intensity is increased not only at the fiber end-tip but also along the z axis.

In conclusion, we investigated light focusing in random media by an adaptive technique in the presence of transverse Anderson localization. By applying an iterative optimization process we were able to feed a localized mode by a spatially modulated beam. The quality of the obtained focus differs strongly from what reported in 3D disordered structures and in standard fibers, because light couples to localized modes traveling for centimeters. We found that the inhibition of diffusion imposed by the Anderson regime boosts the amount of light coupled to the target, lowering the detrimental contribution to the focusing due to the background speckle pattern by several orders of magnitude. The transversal localization hence non-trivially cooperate in the focusing action, and two mechanisms are found to occur: on one hand, the TL located at the target point transports most of the energy, on the other hand, the other modes interfere locally and enhance the intensity.
The resulting focusing efficiency is found to be nearly independent of the fiber length, and this may ultimately lead to a variety of applications based on the transport and focusing of light in specific points also located in distant positions within disordered matter.

\emph{Acknowledgements} S.K. and A.M. are supported by grant number 1029547 from the National Science Foundation.


%

\end{document}